# Quantitative Formulation of Frequency-Dependent Average Force in AM-AFM

Kenichi Umeda[1,2], Karen Kamoshita[3], and Noriyuki Kodera[1]

[1] *Nano Life Science Institute (WPI-NanoLSI),*

      *Kanazawa University, Kakuma-machi, Kanazawa, Ishikawa, 920-1192, Japan.*

[2] *PRESTO/JST, 4-1-8 Honcho, Kawaguchi, Saitama 332-0012, Japan.*

[3] *School of Mathematics and Physics, College of Science and Engineering,*

      *Kanazawa University, Kakuma-machi, Kanazawa, Ishikawa, 920-1192, Japan.*

Corresponding Authors

Dr. Kenichi Umeda (E-mail: umeda.k@staff.kanazawa-u.ac.jp)

Prof. Noriyuki Kodera (E-mail: nkodera@staff.kanazawa-u.ac.jp)

## Abstract

Amplitude-modulation atomic force microscopy (AM-AFM) measures nanoscale surface structures by detecting changes in the cantilever oscillation amplitude, contributing to materials research. AM-AFM can non-destructively observe fragile molecules, such as biomolecules, even while the probe is in intermittent contact with the sample. However, it remains unclear why the tip–sample interaction force estimated from an experimental amplitude value is substantially greater than the actual molecular binding force, despite the successful visualization of molecular dynamics. Here, we formulate a quantitative force conversion equation for arbitrary driving frequencies. Comprehensive theoretical analysis reveals that when the cantilever is excited at the resonance slope, the conventional equation overestimates the actual force by approximately five times, as it is valid only for excitation at the resonance frequency. The theory is validated by simulations and experiments and can be applied to various AM-AFM applications in materials research.

## 1. Introduction

Amplitude-modulation (tapping-mode) atomic force microscopy (AM-AFM), a type of dynamic-mode AFM, is a technique that enables imaging at the atomic and molecular scale with minimal perturbation [1-3]. It has been employed in various materials science research areas, e.g., electrochemistry [4], electronics [5-7], physicochemistry [8], and polymers [5,6,9,10], as it operates effectively in liquid as well as ambient environments. Notably, this technique has contributed to the visualization of living biological samples at the nanoscale [2,5,11-14]. Recent advancements in the bandwidth of AM-AFM devices have facilitated the development of high-speed AFM (HS-AFM), which has uncovered the submolecular-scale dynamics of biomolecules [12,15-21].

A unique feature of AFM is that the tip−sample interaction force can serve not only as a feedback signal but also as a force sensor or stimulator. In bio-applications, it has been employed for the mechanical investigations of force-sensing ion channels [12], and for the manipulation or disassembly of supramolecular complexes [11,13,16-18]. Therefore, quantitative estimation of the tip−sample force is essential for gaining deeper insight into their mechanical properties and contributes to optimizing feedback parameters required for non-destructive imaging of fragile molecules.

There are two primary ways to quantify the forces detected by the AFM tip: the average force and the peak force [22,23]. Most previous research has focused on recovering the peak force, employing force volume mapping [8,10,24-26] or other novel methods [18,22,23,27-29], despite these being both theoretically and experimentally more challenging than estimating the average force. This emphasis is due to the peak force being crucial for evaluating the viscoelastic, chemical, and electronic properties of the sample surface [8,29,30]. However, the peak force is typically estimated to be one or two orders of magnitude greater than the average force [22,23]. When using the previously proposed conversion formula [31,32], the average force is estimated to be about $50-100$ pN under typical experimental conditions for biomolecules; therefore, the peak force reaches the nanonewton range. In contrast, while the binding force of biomolecules typically increases with the loading rate, the equilibrium binding force of fragile biomolecular conjugates can be as small as a few pN [33,34]. Therefore, if the peak force acts effectively on these molecules, it becomes difficult to explain how functional dynamics can be visualized without damaging the molecular structure due to excessive force from the tip.

This discrepancy indicates that the average force mainly contributes the destruction of biomolecules, as the relaxation time of the denaturation of typical biomolecules ($30-500$ μs) [35] is considerably longer than the duration of the peak force (approximately $0.1-1$ μs, estimated from a resonance frequency of $1-0.1$ MHz). This is further supported by single-molecule pulling studies, which have demonstrated that the rupture force increases proportionally with the loading rate [36]. However, even if the average force is effective, it is still estimated to be several times higher than the actual binding force.

Furthermore, in AM-AFM, it is widely recognized that excitation at the resonance slope rather than at the resonance frequency ($f_0$) is ideal for gentle imaging [1]. Although there are examples of studies that have investigated the excitation frequency dependence of forces [23,37-39], their findings have been limited to partial insights, and comprehensive research with sufficient experimental and simulation-based validation has yet to be conducted.

In this study, we derive a quantitative conversion equation to estimate the average force from the amplitude values at arbitrary drive frequencies. Our comprehensive theoretical analysis reveals that the estimated average force is highly dependent on the driving frequency and can be up to five times smaller than when excited at $f_0$, thereby resolving one of the longstanding enigmas in the field of AM-AFM. The theoretical framework presented here is anticipated to have a significant impact on various materials research fields, particularly studies involving fragile materials, including biological samples, where minimizing applied force is crucial for accurate observations.

## 2. Conversion of Amplitude to Average Force

In AM-AFM, to observe fragile biomolecules without destroying them, it is generally recognized that driving the cantilever at the resonance slope is essential to achieve the maximum force sensitivity [1]. However, most of previous studies have focused on the excitation at $f_0$ and there is no comprehensive theoretical research that can be applied to any given frequency. Additionally, compared to the peak force, the average force can be more easily estimated from the amplitude setpoint value without the need for time-consuming experiments [31,32], making it more compatible with imaging fragile biomolecules in action. Therefore, in this section, we derive a theoretical equation for the average force that can be applied to an arbitrary driving frequency ($f_{drive}$).

The equation of motion of the cantilever based on a simple harmonic oscillator (SHO) model is expressed as follows:

$$\frac{k_{cl}}{\omega_0^2} z''_{tip}(t) + \frac{k_{cl}}{\omega_0 Q_{cl}} z'_{tip}(t) + k_{cl} z_{tip}(t) = \underbrace{F_{drive} \cos(\omega_{drive} t)}_{\text{External driving force}} + \underbrace{F_{ts}\left(z_{tip}(t), z'_{tip}(t)\right)}_{\text{Tip-sample force}}, \qquad (1)$$

where $\omega_0$ (= $2\pi f_0$), $Q_{cl}$, $k_{cl}$, $F_{drive}$, $\omega_{drive}$ (= $2\pi f_{drive}$), and $z_{tip}$ are the eigenfrequency, quality factor, spring constant, driving force, driving frequency, and displacement of the cantilever, respectively. Hereafter, we denote $f$ and $\omega$ (= $2\pi f$) as a certain frequency and its angular frequency, respectively.

Normally, the distance dependence of the tip–sample interaction is nonlinear; however, when the interaction depth is relatively small, the harmonic components can be ignored. Assuming a steady-state oscillation, $z_{tip}$ can be approximately expressed as

$$z_{tip}(t) \approx \left\langle z_{tip} \right\rangle + A_{cl} \cos\left(\omega_{drive} t + \phi_{cl}\right), \qquad (2)$$

where $A_{cl}$ and $\phi_{cl}$ are the oscillation amplitude and phase delay in the excitation system, respectively. Using the Fourier series of the tip–sample interaction, we solve this equation to obtain the transfer function $G_{cl}$ as follows [26]:

$$G_{cl}(\tilde{\omega}_{drive}) = \frac{1}{k_{cl}} \frac{1}{\left(1 - \tilde{\omega}_{drive}^2 - I_{even}\left(z_{tip}, A_{cl}\right)\right) + i\left(\tilde{\omega}_{drive} / Q_{cl} + I_{odd}\left(z_{tip}, A_{cl}\right)\right)}, \qquad (3)$$

where $\tilde{\omega}_{drive}$ is a normalized $f_{drive}$ that is defined as follows:

$$\tilde{\omega}_{drive} \equiv \frac{\omega_{drive}}{\omega_0} = \frac{f_{drive}}{f_0}, \qquad (4)$$

and $I_{even}$ and $I_{odd}$ represent the in-phase and out-of-phase terms of the Fourier components, reflecting conservative forces and energy dissipation, respectively [26], and are expressed as,

$$\begin{aligned} I_{even}(z_{tip}, A_{cl}) &= \frac{2 f_{drive}}{k_{cl} A_{cl}} \int_0^{1/f_{drive}} F_{ts}\left(z_{tip}(t), z'_{tip}(t)\right) \cos\left(\omega_{drive} t + \phi_{drive}\right) dt \\ &= \frac{2}{k_{cl} A_{cl}^2} \left\langle F_{ts} \cdot z_{tip} \right\rangle \propto \Delta f, \end{aligned} \qquad (5)$$

$$\begin{aligned} I_{odd}(z_{tip}, A_{cl}) &= \frac{2 f_{drive}}{k_{cl} A_{cl}} \int_0^{1/f_{drive}} F_{ts}\left(z_{tip}(t), z'_{tip}(t)\right) \sin\left(\omega_{drive} t + \phi_{drive}\right) dt \\ &= -\frac{1}{\pi f_{drive} k_{cl} A_{cl}^2} \left\langle F_{ts} \cdot z'_{tip} \right\rangle, \end{aligned} \qquad (6)$$

where $F_{ts}$ is the tip–sample interaction force. By following the original definition by Dürig et al. [40], we define them as even or odd relative to the tip trajectory. Moreover, $A_{cl}$ and $\phi_{cl}$ are calculated as follows:

$$\begin{aligned} A_{cl}(\tilde{\omega}_{drive}) &= \left| G_{cl}(\tilde{\omega}_{drive}) \right| F_{drive} \\ &= \frac{A_0}{\sqrt{\left(1 - \tilde{\omega}_{drive}^2 - I_{even}\left(z_{tip}, A_{cl}\right)\right)^2 + \left(\tilde{\omega}_{drive} / Q_{cl} + I_{odd}\left(z_{tip}, A_{cl}\right)\right)^2}}, \end{aligned} \qquad (7)$$

$$\phi_{cl}(\tilde{\omega}_{drive}) = \angle G_{cl}(\tilde{\omega}_{drive}) = \tan^{-1}\left(-\frac{\tilde{\omega}_{drive} / Q_{cl} + I_{odd}\left(z_{tip}, A_{cl}\right)}{1 - \tilde{\omega}_{drive}^2 - I_{even}\left(z_{tip}, A_{cl}\right)}\right), \qquad (8)$$

respectively, where $A_0$ represents $A_{cl}$ at $\tilde{\omega}_{drive} = 0$. When $F_{ts} = 0$, $A_{cl}$ is expressed by

$$A_{cl}(\tilde{\omega}_{drive}, I_{even} = I_{odd} = 0) = \frac{A_0}{\sqrt{\left(1 - \tilde{\omega}_{drive}^2\right)^2 + \left(\tilde{\omega}_{drive} / Q_{cl}\right)^2}}, \qquad (9)$$

The frequency characteristics of amplitude as a function of $I_{even}$ are analyzed in Fig. 1(a). As the repulsive force is increased ($I_{even} < 0$), a positive $f_0$ shift and a reduction in the overall excitation efficiency are observed, which can be attributed to an increase in the effective $k_{cl}$. In addition, the resonance peak becomes slightly sharper, meaning that the effective $Q_{cl}$ increases, which makes the amplitude change at the resonance peak less sensitive to the force. In contrast, as the attractive force is increased ($I_{even} > 0$), a negative frequency shift and an increase in the overall excitation efficiency are observed due to the reduction of the effective $k_{cl}$.

Moreover, the frequency characteristics of amplitude as a function of $I_{odd}$ are analyzed in Fig. 1(b). As $I_{odd}$ increases, $Q_{cl}$ and the overall excitation efficiency decreases, but the $f_0$ shift is not observed. In other words, the maximum sensitivity for $I_{odd}$ can be obtained by exciting at the frequency near the resonance peak (more detailed analysis of dissipation will be published elsewhere).

Enlarged views of the frequency characteristics near the resonance are shown in Fig. 1(c,d). It is well known that when $Q_{cl}$ is sufficiently large, the maximum amplitude appears exactly at $f_0$. However, as seen here, when $Q_{cl}$ becomes significantly low, the resonance peak frequency ($f_{peak}$) deviates slightly negative from $f_0$. The analytical solution for $f_{peak}$ can be derived by setting the derivative of $A_{cl}$ in Eq. (9) to zero, as follows:

$$\frac{\partial A_{cl}(\tilde{\omega}_{drive})}{\partial \tilde{\omega}_{drive}} = A_0 \frac{\tilde{\omega}_{drive}\left[2\left(1-\tilde{\omega}_{drive}^2\right)-1/Q_{cl}^2\right]}{\left[\left(1-\tilde{\omega}_{drive}^2\right)^2+\left(\tilde{\omega}_{drive}/Q_{cl}\right)^2\right]^{3/2}} = 0. \tag{10}$$

Solving this equation, we obtain the expression of normalized $f_{peak}$ ($\tilde{\omega}_{peak}$) as follows:

$$\tilde{\omega}_{peak} \equiv \frac{\omega_{peak}}{\omega_0} = \frac{f_{peak}}{f_0} = \sqrt{1-\frac{1}{2Q_{cl}^2}} \quad \left(Q_{cl} > \frac{1}{\sqrt{2}}\right). \tag{11}$$

In Fig. 1(c), when excited at $f_{peak}$ or lower, the amplitude value tends to effectively decrease with reducing $I_{even}$. However, even when $I_{even}$ is changed, the amplitude value is almost unchanged at $f_0$ and rather increases at a frequency higher than $f_0$. In contrast, in Fig. 1(d), as $I_{odd}$ is changed, the amplitude value decreases regardless of $f_{drive}$, especially near $f_{peak}$.

As AM-AFM is also known as the slope detection method [1,3,41], it has been theoretically clarified that the sensitivity of conservative forces, which are reflected in the $f_0$ shift, is also maximized when the cantilever is excited at the MaxSlope frequency ($f_{MS}$) at which the amplitude slope is at its maximum. Under ambient conditions [7], the upper MaxSlope frequency ($f_{UMS}$ in Fig. 1(c,d)) is typically excited to effectively detect the negative frequency shift resulting from attractive interactions, whereas in liquid environments, the lower MaxSlope frequency ($f_{LMS}$ in Fig. 1(c,d)) is excited to effectively detect the positive frequency shift resulting from repulsive interactions. If the attractive and repulsive forces are switched during imaging, the amplitude would not decrease monotonically as the tip approaches, leading to unstable imaging [26,42,43]. Particularly, in liquids, excitation at a slightly lower off-resonance frequency is important not only to improve the sensitivity, but also to compensate for the slight negative shift of $f_0$ caused by the squeeze film effect during approach [44]. For these reasons, excitation at the slightly off-resonance frequency is typically recommended in most AM-AFM operation manuals [45-49].

In HS-AFM experiments, most protein surfaces are relatively elastic and imaging is performed with the minimum force required; therefore, $I_{odd}$ can be negligible compared to $I_{even}$. Furthermore, in Eq. (7), by using the free oscillation amplitude ($A_{free}$), $A_{cl}$ can be expressed as follows:

$$A_{cl}(\tilde{\omega}_{drive}) = A_{free} \sqrt{\frac{\left(1-\tilde{\omega}_{drive}^2\right)^2 + \left(\tilde{\omega}_{drive}/Q_{cl}\right)^2}{\left(1-\tilde{\omega}_{drive}^2 - \dfrac{2}{k_{cl}A_{cl}^2}\left\langle F_{ts}\cdot z_{tip}\right\rangle\right)^2 + \left(\tilde{\omega}_{drive}/Q_{cl}\right)^2}}. \tag{12}$$

This equation is an implicit function of $A_{cl}$ because $A_{cl}$ appears on both sides. When the oscillation amplitude is sufficiently large, such that the tip is in contact with the sample for only a short time during a cantilever period, the large-amplitude approximation can be applied as follows [42]:

$$\lim_{A_{free}\to\infty} \left\langle F_{ts}\cdot z_{tip}\right\rangle \approx -A_{cl}\left\langle F_{ts}\right\rangle. \tag{13}$$

By solving Eq. (12) for $\left\langle F_{ts}\right\rangle$, we obtain the force conversion equation at arbitrary $f_{drive}$ as follows:

$$\left\langle F_{ts}\right\rangle = \frac{k_{cl}A_{free}}{2}\left[-\left(1-\tilde{\omega}_{drive}^2\right)\tilde{A}_{cl} \overset{rep}{\underset{att}{\pm}} \sqrt{\left(1-\tilde{\omega}_{drive}^2\right)^2 + \left(\tilde{\omega}_{drive}/Q_{cl}\right)^2\left(1-\tilde{A}_{cl}^2\right)}\right], \tag{14}$$

where the positive and negative signs are used for the repulsive ($\left\langle F_{ts}\right\rangle > 0$) and attractive ($\left\langle F_{ts}\right\rangle < 0$) regimes, respectively, and $\tilde{A}_{cl}$ is a normalized amplitude that is defined as follows:

$$\tilde{A}_{cl} \equiv \frac{A_{cl}}{A_{free}} = \frac{A_{free} + \Delta A_{ts}}{A_{free}}, \tag{15}$$

where $\Delta A_{ts}$ is the amplitude change resulting from the tip–sample interaction. When the feedback error is neglected, $\tilde{A}_{cl}$ represents the setpoint ratio. By substituting $\omega_0$ for $\omega_{drive}$, namely, $\tilde{\omega}_{drive} = 1$, we obtain a well-known conventional conversion equation as follows [9,12,15,17,32]:

$$\langle F_{ts} \rangle = \frac{k_{cl}}{2Q_{cl}} \sqrt{A_{free}^2 - A_{cl}^2} = \frac{k_{cl} A_{free}}{2Q_{cl}} \sqrt{1 - \tilde{A}_{cl}}. \tag{16}$$

Conversely, the equation for calculating $A_{cl}$ from $\langle F_{ts} \rangle$ is obtained by solving Eq. (12) for $A_{cl}$ as follows:

$$A_{cl}(\tilde{\omega}_{drive}) = \sqrt{A_{free}^2 - (\tilde{\omega}_{drive} / Q_{cl})^2 \beta^2} - (1 - \tilde{\omega}_{drive}^2)\beta, \tag{17}$$

where

$$\beta = \frac{2}{\left[\left(1 - \tilde{\omega}_{drive}^2\right)^2 + \left(\tilde{\omega}_{drive} / Q_{cl}\right)^2\right] k_{cl}} \langle F_{ts} \rangle. \tag{18}$$

Substituting $\tilde{\omega}_{drive} = 1$ yields another common form as follows [14,31]:

$$A_{cl} = A_{free} \sqrt{1 - 4\left(\frac{\langle F_{ts} \rangle}{F_{drive}\big|_{\tilde{\omega}=1}}\right)^2}, \tag{19}$$

where $F_{drive}$ is given by

$$F_{drive}\big|_{\tilde{\omega}=1} = \frac{A_{free} k_{cl}}{Q_{cl}}. \tag{20}$$

It is thus demonstrated that the equation derived here for arbitrary $f_{drive}$ includes the equations previously derived for $f_0$.

## 3. Driving Frequency Dependence of Forces

In this section, we examine the frequency characteristics of the derived formula. First, in the repulsive regime, Fig. 2(a) presents the correlation between $\tilde{A}_{cl}$ and the converted force at various $f_{drive}$. The calculation was performed using a typical experimental condition of HS-AFM: $k_{cl} = 0.1$ N/m, $f_0 = 1$ MHz, $Q_{cl} = 1.5$, and $A_{free} = 3$ nm$_{p-0}$. When $f_{drive}$ is set to $f_0$, the curve is nonlinear, and the force rises steeply near $\tilde{A}_{cl} = 1$. In this condition, even a slight change in the amplitude setpoint results in a significant variation in the force, making it difficult to measure accurately with a small force. In contrast, as $f_{drive}$ is decreased, the force slope becomes smaller and closer to linear, making it possible to measure with a small force. Near the MaxSlope frequency, the force slope is minimized and becomes almost linear. However, as $f_{drive}$ is further decreased, the slope increases once again, leading to a deterioration in force sensitivity.

Since excitation at $f_{peak}$ is also commonly employed, we also calculate the force when excited at $f_{peak}$. Consequently, the force slope increases to ~1.5 times greater that at the MaxSlope frequency; however, it remains considerably smaller than that for excitation at $f_0$. This result clearly indicates that the conventional force conversion method overestimates the applied force by several times compared to the actual force when excited at either $f_{peak}$ or the MaxSlope frequencies.

Furthermore, in Fig. 2(b), we also quantified the $f_{drive}$ dependence of the force (normalized by those at $f_0$) at $\tilde{A}_{cl} = 0.95$. We observed that force minima occur at a frequency lower than $f_0$ (see the arrows), which is referred to as the MinForce frequency ($f_{MF}$). As is the case for MaxSlope, MinForce exists at both lower and upper frequencies, denoted as $f_{LMF}$ and $f_{UMF}$, respectively. We also found that the prominence of this minimum increases as $Q_{cl}$ increases. When $Q_{cl} = 1.5$, this minimum force is 1/5 of the force at $f_0$. This result suggests that excitation at the MinForce frequency enables imaging with minimal force.

For example, the conventional equation for $f_0$ estimates the interaction force at $\tilde{A}_{cl} = 0.95$ to be 31 pN. This force is substantially stronger than the typical biological binding force of ~10 pN; e.g., the binding force between myosin and F-actin is 15 pN [33]. In contrast, our improved equation estimates the interaction force at $\tilde{A}_{cl} = 0.95$ to be 6.7 pN when excited at MaxSlope frequency. This force level is below the strength of typical intermolecular bonds, demonstrating that this provides a consistent explanation for the force overestimation problem.

For a more comprehensive understanding, in Fig. 2(c), we also conduct the calculation at frequencies exceeding $f_0$. The results display the characteristics of a two-valued curve, where the branch in the far region can be calculated by applying the negative sign in Eq. (14). This indicates that as the tip approaches the surface, the amplitude undergoes an initial increase until $f_0$ exceeds $f_{drive}$, followed by a subsequent decline, as predicted from the results in Fig. 1(a,c). This causes an excessive force being applied, even if $\tilde{A}_{cl} \sim 1$. Occasionally, this leads to the undesirable inversion of image contrast, where a lower area is observed as higher than a higher area, distorting the imaging interpretation. Note that there are also important precautions to consider when experimentally converting the amplitude value into force (Supporting Information 1).

We next derive an analytical expression for the MinForce frequencies. To obtain this, the force magnitude near $\tilde{A}_{cl} = 1$ in Fig. 2(a) must be examined; however it drops to zero, making a direct comparison impossible. Instead, when excited below the resonance slope frequencies, the force increases linearly, particularly at $\tilde{A}_{cl} \sim 1$, and is approximated by

$$\langle F_{ts} \rangle \approx \left( \frac{1}{A_{free}} \lim_{A_{cl} \to 1} \frac{\partial \langle F_{ts} \rangle}{\partial \tilde{A}_{cl}} \right) \Delta A_{ts} \equiv \alpha_{\Delta A \to F} \Delta A_{ts}. \tag{21}$$

This equation indicates that the force magnitude is correlated with the force slope near $\tilde{A}_{cl} = 1$. Therefore, we obtain the force slope by differentiating Eq. (14) as follows:

$$\frac{\partial \langle F_{ts} \rangle}{\partial \tilde{A}_{cl}} = -\frac{k_{cl} A_{free}}{2Q_{cl}} \left[ \frac{\tilde{A}_{cl} \tilde{\omega}_{drive}^2}{\sqrt{\left(1 - \tilde{A}_{cl}^2\right) \tilde{\omega}_{drive}^2 + Q_{cl}^2 \left(1 - \tilde{\omega}_{drive}^2\right)^2}} + Q_{cl} \left(1 - \tilde{\omega}_{drive}^2\right) \right]. \tag{22}$$

Then, $\alpha_{\Delta A \to F}$, which represents the conversion coefficient from $\Delta A_{ts}$ to $\langle F_{ts} \rangle$ at $\tilde{A}_{cl} = 1$, is obtained as follows:

$$\begin{aligned}
\alpha_{\Delta A \to F} &\equiv \frac{1}{A_{free}} \lim_{\tilde{A}_{cl} \to 1} \frac{\partial \langle F_{ts} \rangle}{\partial \tilde{A}_{cl}} \\
&= -\frac{k_z}{2Q_{cl}} \left[ \frac{\tilde{\omega}_{drive}^2}{Q_{cl} \left(1 - \tilde{\omega}_{drive}^2\right)} + Q_{cl} \left(1 - \tilde{\omega}_{drive}^2\right) \right] \quad \left(\tilde{\omega}_{drive} \neq 1\right).
\end{aligned} \tag{23}$$

This equation is plotted in Fig. 2(d), which bears a close resemblance to that in Fig. 2(c). Note that Eqs. (22) and (23) are valid only for $\tilde{\omega}_{drive} < 1$ and $\tilde{\omega}_{drive} > 1$ in the repulsive and attractive regimes, respectively. By solving $\partial \alpha_{\Delta A \to F} / \partial \tilde{\omega}_{drive} = 0$, the MinForce frequencies are obtained as follows (see the arrows):

$$\tilde{\omega}_{LMF} = \frac{f_{LMF}}{f_0} = \sqrt{1 - \frac{1}{Q_{cl}}} \quad \left(Q_{cl} \geq 1\right), \tag{24}$$

$$\tilde{\omega}_{UMF} = \frac{f_{UMF}}{f_0} = \sqrt{1 + \frac{1}{Q_{cl}}}. \tag{25}$$

Importantly, this analysis revealed that the MinForce frequencies can be expressed by straightforward analytical equations.

Previous studies have reported an equation describing the frequency dependence of peak force based on the Hertzian model approximation [23,39]. While this equation enables the calculation of peak force without time-consuming experiments, its applicability is limited to simple geometries where the Hertzian theory holds. Additionally, various parameters such as the Young's modulus and shape of both the tip and the sample must be assumed. Using the same frequency analysis, we found that the MinForce frequencies of the peak force are also given by Eqs. (24) and (25) (Supporting Information 2). Thus, excitation at the MinForce frequency maximizes force detection sensitivity for both the average and peak forces.

A previous study empirically derived a linear approximation for force conversion at off-resonance frequencies based on simulation results, as follows [37]:

$$\left\langle F_{\mathrm{ts}} \right\rangle_{\mathrm{off\text{-}reso}} = -\frac{k_{\mathrm{cl}}}{2Q_{\mathrm{eff}}} \Delta A_{\mathrm{ts}},$$ (26)

where $Q_{\mathrm{eff}}$ denotes the effective $Q_{\mathrm{cl}}$, which can be derived from Eq. (9) as follows:

$$Q_{\mathrm{eff}} = \frac{1}{\sqrt{\left(1 - \tilde{\omega}_{\mathrm{drive}}^2\right)^2 + \left(\tilde{\omega}_{\mathrm{drive}} / Q_{\mathrm{cl}}\right)^2}},$$ (27)

When $\tilde{\omega}_{\mathrm{drive}} = 1$, $Q_{\mathrm{eff}}$ equals $Q_{\mathrm{cl}}$, whereas when $\tilde{\omega}_{\mathrm{drive}} = 0$, $Q_{\mathrm{eff}}$ decreases to 1.

In the following steps, we demonstrate that Eq. (26) can be derived from Eq. (23). By solving Eq. (27) for $\tilde{\omega}_{\mathrm{drive}}$, the following expression is obtained:

$$\tilde{\omega}_{\mathrm{drive}} = \sqrt{1 - \frac{1}{2Q_{\mathrm{cl}}^2} - \sqrt{\frac{1}{4Q_{\mathrm{cl}}^4} - \frac{1}{Q_{\mathrm{cl}}^2} + \frac{1}{Q_{\mathrm{eff}}^2}}}.$$ (28)

When $Q_{\mathrm{cl}}$ is sufficiently large (typically $Q_{\mathrm{cl}} > 10$), it can be approximated as follows:

$$\lim_{Q_{\mathrm{cl}} \to \infty} \tilde{\omega}_{\mathrm{drive}} \approx \sqrt{1 - \frac{1}{Q_{\mathrm{eff}}}}.$$ (29)

Substituting this into Eq. (23) yields:

$$\alpha_{\Delta A \to F}\big|_{\mathrm{off\text{-}reso}} = -\frac{k_{\mathrm{cl}}}{2Q_{\mathrm{cl}}} \left( \frac{Q_{\mathrm{eff}}}{Q_{\mathrm{cl}}} + \frac{Q_{\mathrm{cl}}}{Q_{\mathrm{eff}}} - \frac{1}{Q_{\mathrm{cl}}} \right).$$ (30)

Again, by taking the limit as $Q_{\mathrm{cl}} \to \infty$, this equation can be simplified as follows:

$$\lim_{Q_{\mathrm{cl}} \to \infty} \alpha_{\Delta A \to F}\big|_{\mathrm{off\text{-}reso}} = -\frac{k_{\mathrm{cl}}}{2Q_{\mathrm{eff}}}.$$ (31)

Substituting this into Eq. (21) yields Eq. (26), which provides good approximation when $f_{\mathrm{drive}}$ is smaller than the MinForce frequency, indicating that Eq. (23) encompasses Eq. (26).

Next, the attractive regime is also analyzed. Since it is only utilized under ambient environments, typical values for ambient AM-AFM of $Q_{\mathrm{cl}} = 300$ and $k_{\mathrm{cl}} = 3$ N/m were used. In Fig. 2(e), we then obtained results showing nearly the same tendency as the repulsive force presented in Fig. 2(a). When excited at $f_0$, the force increases steeply at $\tilde{A}_{cl} = 1$, but as $f_{drive}$ increases, the force slope becomes smaller and approaches linear. When $f_{drive}$ exceeds the upper MinForce frequency, the force slope increases again. In Fig. 2(f), we also examine the $\tilde{\omega}_{drive}$ dependence of the force slope and obtained results similar to those for the repulsive regime in Fig. 2(d), except that the curves are inverted along the X-axis. Another notable difference is that while the values converge to a constant at low frequencies in the repulsive regime, they diverge at high frequencies in the attractive regime.

Substituting Eqs. (24) and (25) into Eq. (14) gives simplified conversion equations from amplitude value to force at MinForce as follows:

$$\left\langle F_{ts} \right\rangle_{LMF} = \frac{k_{cl} A_{free}}{2 Q_{cl}} \left[ -\tilde{A}_{cl} + \sqrt{1 + \left(1 - \frac{1}{Q_{cl}}\right)\left(1 - \tilde{A}_{cl}^2\right)} \right] \quad (Q_{cl} \geq 1), \tag{32}$$

$$\left\langle F_{ts} \right\rangle_{UMF} = -\frac{k_{cl} A_{free}}{2 Q_{cl}} \left[ -\tilde{A}_{cl} + \sqrt{1 + \left(1 + \frac{1}{Q_{cl}}\right)\left(1 - \tilde{A}_{cl}^2\right)} \right]. \tag{33}$$

By calculating the derivative according to Eq. (23), $\alpha_{\Delta A \to F}$ are obtained as follows:

$$\alpha_{\Delta A \to F}(\tilde{\omega}_{LMF}) = -\frac{k_{cl}}{2 Q_{cl}} \left(2 - \frac{1}{Q_{cl}}\right) \quad \text{if } \left\langle F_{ts} \right\rangle > 0, \tag{34}$$

$$\alpha_{\Delta A \to F}(\tilde{\omega}_{UMF}) = \frac{k_{cl}}{2 Q_{cl}} \left(2 + \frac{1}{Q_{cl}}\right) \quad \text{if } \left\langle F_{ts} \right\rangle < 0. \tag{35}$$

These equations represent the force conversion at the frequency where the force detection sensitivity is maximized.

However, since excitation is conventionally performed at the MaxSlope frequency [1], we next need to identify the frequency offering the greatest advantage. To predict the MaxSlope frequency, we must solve the second-order differential equation of $A_{cl}$ (Eq. (9)) as follows:

$$\frac{\partial^2 A_{cl}(\tilde{\omega}_{drive})}{\partial \tilde{\omega}_{drive}^2} = A_0 \left\{ \frac{3\tilde{\omega}_{drive}^2 \left[2\left(1 - \tilde{\omega}_{drive}^2\right) - 1/Q_{cl}^2\right]^2}{\left[\left(1 - \tilde{\omega}_{drive}^2\right)^2 + \left(\tilde{\omega}_{drive}/Q_{cl}\right)^2\right]^{5/2}} + \frac{2\left(1 - \tilde{\omega}_{drive}^2\right) - 1/Q_{cl}^2 - 4\tilde{\omega}_{drive}^2}{\left[\left(1 - \tilde{\omega}_{drive}^2\right)^2 + \left(\tilde{\omega}_{drive}/Q_{cl}\right)^2\right]^{3/2}} \right\}$$
$$= 0.$$

$$(36)$$

Unfortunately, this equation is difficult to solve by elementary functions. However, when $Q_{cl}$ is sufficiently large ($Q_{cl} \gg 10$), $A_{cl}$ in Eq. (9) can be approximated by a Lorentzian function as follows:

$$\lim_{Q_{cl} \to \infty} A_{cl}(\tilde{\omega}_{drive}) = \frac{A_0}{\sqrt{4\left(1 - \tilde{\omega}_{drive}\right)^2 + 1/Q_{cl}^2}},$$ (37)

which is given by a Taylor series approximation. With this equation, the differential equation above can be analytically solved as follows [1]:

$$\lim_{Q_{cl} \to \infty} \tilde{\omega}_{LMS} = \lim_{Q_{cl} \to \infty} \frac{f_{LMS}}{f_0} \approx 1 - \frac{1}{\sqrt{8}Q_{cl}},$$ (38)

$$\lim_{Q_{cl} \to \infty} \tilde{\omega}_{UMS} = \lim_{Q_{cl} \to \infty} \frac{f_{UMS}}{f_0} \approx 1 + \frac{1}{\sqrt{8}Q_{cl}}.$$ (39)

This approximation cannot be applied to liquid AM-AFM, where $Q_{cl}$ is substantially reduced (1.5 for typical HS-AFM experiments). Therefore, by expanding these approximate solutions with the numerically calculated solution of Eq. (36), we obtained a Laurent polynomial approximation for the lower MaxSlope ($\tilde{\omega}_{LMS}$) as follows:

$$\tilde{\omega}_{LMS} \approx 1 - \frac{1}{Q_{cl}}\left[\frac{1}{\sqrt{8}} + \frac{0.2148}{Q_{1/2}} - \left(\frac{0.3074}{Q_{1/2}}\right)^2 + \left(\frac{0.2739}{Q_{1/2}}\right)^3 - \left(\frac{0.2016}{Q_{1/2}}\right)^4 + \left(\frac{0.1715}{Q_{1/2}}\right)^{12}\right],$$

where $Q_{1/2} = Q_{cl} - 1/2 \quad \left(Q_{cl} \geq 1/\sqrt{2}\right),$

$$(40)$$

which is valid for arbitrary $Q_{cl}$. When $Q_{cl} = 1/\sqrt{2}$, $\tilde{\omega}_{LMS}$ goes down to zero. In contrast, the $Q_{cl}$ dependency of the upper MaxSlope ($\tilde{\omega}_{UMS}$) forms a biphasic curve; therefore, by switching two equations between high and low $Q_{cl}$, it can be analytically expressed over the entire $Q_{cl}$ range (Supporting Note 3). The equation for high $Q_{cl}$, which is significant in experiments, is given as follows:

$$\tilde{\omega}_{UMS} \approx 1 + \frac{1}{Q_{cl}}\left[\frac{1}{\sqrt{8}} - \frac{0.2194}{Q_{1/4}} + \left(\frac{0.2294}{Q_{1/4}}\right)^2 - \left(\frac{0.2547}{Q_{1/4}}\right)^3 + \left(\frac{0.2131}{Q_{1/4}}\right)^4\right],$$

where $Q_{1/4} = Q_{cl} - 1/4,$

$$(41)$$

which is valid for arbitrary $Q_{cl}$ greater than 0.5.

We next compare the $Q_{cl}$ dependence of the MinForce and MaxSlope frequencies in the repulsive regime. In Fig. 3(a), as $Q_{cl}$ decreases, the lower MinForce frequency steeply falls down to 0 even at $Q_{cl} = 1$ while the MaxSlope frequency remains above 0.4 at $Q_{cl} > 1$. Moreover, the forces normalized by those at $f_0$ are compared in Fig. 3(b), exhibiting that there is almost no significant difference across the entire $Q_{cl}$ value. It can be seen that the force at $f_{peak}$ is several times larger than them. For ease of understanding, the values for a typical HS-AFM parameter ($Q_{cl} = 1.5$) are also shown in Fig. 3(a,b). A notable discrepancy is observed in the frequencies, with a difference of 0.09 ($= 0.67 − 0.58$) between the MaxSlope and MinForce. In contrast, the difference in forces is relatively minor, at 0.01 ($= 0.22 − 0.21$). This illustrates a significant contrast between the two variables.

We also performed the same analysis in the attractive regime. In Fig. 3(c), the result clearly indicates that the upper MinForce frequency is higher than the upper MaxSlope frequency at all $Q_{cl}$. As $Q_{cl}$ decreases, the upper MinForce frequency increases monotonically while the upper MaxSlope frequency increases once and then decreases. In Fig. 3(d), when analyzing the forces normalized by those at $f_0$, contrary to the repulsive regime, there is a distinct difference between the two forces at $Q_{cl} < 4$. However, as $Q_{cl}$ increases, both the forces asymptotically approach the same value, similar to the repulsive regime.

To compare the two forces more quantitatively, the $Q_{cl}$ dependence of the ratio of $\alpha_{\Delta A \rightarrow F}$ at the MaxSlope and MinForce frequencies is examined in Fig. 3(e). In the repulsive regime, the ratio ranges from 1.03 to 1.06 for all $Q_{cl}$ values, showing no significant $Q_{cl}$ dependence. In the attractive regime, however, the ratio decreases steeply with decreasing $Q_{cl}$. Both the ratios asymptotically approach 1.061, which can be obtained analytically in the following steps. In the large $Q_{cl}$ limit, $\alpha_{\Delta A \rightarrow F}$ for MinForce can be obtained from Eqs. (34) and (35) as follows:

$$\lim_{Q_{cl} \rightarrow \infty} \alpha_{\Delta A \rightarrow F}(\tilde{\omega}_{LMF}) = -\frac{k_{cl}}{Q_{cl}} \quad \text{if } \langle F_{ts} \rangle > 0, \tag{42}$$

$$\lim_{Q_{cl} \rightarrow \infty} \alpha_{\Delta A \rightarrow F}(\tilde{\omega}_{UMF}) = \frac{k_{cl}}{Q_{cl}} \quad \text{if } \langle F_{ts} \rangle < 0. \tag{43}$$

For MaxSlope, these can be obtained by substituting Eqs. (38) and (39) to Eq. (23) as follows:

$$\lim_{Q_{\mathrm{cl}} \to \infty} \alpha_{\Delta A \to F}(\tilde{\omega}_{\mathrm{LMS}}) = -\frac{k_{\mathrm{cl}}}{Q_{\mathrm{cl}}} \frac{3\sqrt{2}}{4} \quad \text{if } \langle F_{\mathrm{ts}} \rangle > 0, \tag{44}$$

$$\lim_{Q_{\mathrm{cl}} \to \infty} \alpha_{\Delta A \to F}(\tilde{\omega}_{\mathrm{UMS}}) = \frac{k_{\mathrm{cl}}}{Q_{\mathrm{cl}}} \frac{3\sqrt{2}}{4} \quad \text{if } \langle F_{\mathrm{ts}} \rangle < 0. \tag{45}$$

Consequently, the value of 1.061 can be obtained as follows:

$$\lim_{Q_{\mathrm{cl}} \to \infty} \frac{\alpha_{\Delta A \to F}(\tilde{\omega}_{\mathrm{LMS}})}{\alpha_{\Delta A \to F}(\tilde{\omega}_{\mathrm{LMF}})} = \lim_{Q_{\mathrm{cl}} \to \infty} \frac{\alpha_{\Delta A \to F}(\tilde{\omega}_{\mathrm{UMS}})}{\alpha_{\Delta A \to F}(\tilde{\omega}_{\mathrm{UMF}})} = \frac{3\sqrt{2}}{4} \approx 1.061, \tag{46}$$

Although the force sensitivity of MinForce is slightly higher than that of MaxSlope, our results indicate that the difference is only ~3% in liquid and ~6% in air, which may fall within the error margin. The results indicate that the approximate equation for MinForce can be employed either when excited at the MaxSlope or MinForce frequency.

To convert amplitude to force, the nonlinear equations presented in Eqs. (14), (32), or (33) are generally required. However, a linear approximation of Eq. (21) can be conveniently applied when exciting at $f_0$, at which $\alpha_{\Delta A \to F}$ can be readily calculated using Eq. (34) or (35). For a typical HS-AFM parameter of $Q_{\mathrm{cl}} = 1.5$, the force can be calculated as follows:

$$\begin{aligned} \langle F_{\mathrm{ts}} \rangle_{\mathrm{LMF}} &\approx -0.444 \cdot k_{\mathrm{cl}} \Delta A_{\mathrm{ts}} \\ &= 0.444 \cdot k_{\mathrm{cl}} A_{\mathrm{free}} \left( 1 - \tilde{A}_{\mathrm{cl}} \right). \end{aligned} \tag{47}$$

By setting other typical values, $k_{\mathrm{cl}} = 0.1$ N/m and $A_{\mathrm{free}} = 3$ nm$_{\mathrm{p-0}}$, a more practical expression can be obtained as follows:

$$\langle F_{\mathrm{ts}} \rangle_{\mathrm{LMF}} \approx 133 \times 10^{-12} \cdot \left( 1 - \tilde{A}_{\mathrm{cl}} \right). \tag{48}$$

Using this equation, for example, we can easily estimate $F_{\mathrm{ts}}$ to be 6.7 and 13.3 pN for $\tilde{A}_{\mathrm{cl}} = 0.95$ and 0.9, respectively. The force of 6.7 pN is consistent with the value of 6.6 pN predicted by the nonlinear equation in Eq. (32).

## 4. Numerical Simulation-Based Validation of Theory

To validate the applicability of the derived theory, we conducted numerical simulations based on the Hertzian theory for contact between a sphere and an infinite plane, as follows [26,50-52]:

$$F_{\text{Hertz}}(\delta) = \frac{4E^*}{3}\sqrt{R_{\text{tip}}}\left(-\delta\right)^{3/2}, \tag{49}$$

where $R_{\text{tip}}$ and $\delta$ are the radius of curvature of the tip and indentation depth, respectively, and $E^*$ is the reduced Young's modulus, expressed as

$$E^* = \left(\frac{1-\nu_s^2}{E_s} + \frac{1-\nu_{\text{tip}}^2}{E_{\text{tip}}}\right)^{-1}, \tag{50}$$

where $\nu$ and $E$ represent the Poisson's ratio and Young's modulus, respectively, where the subscripts s and tip denote the sample and tip, respectively. When the tip is sufficiently rigid compared to the sample, $E^*$ can be approximated as follows:

$$E^* \approx \frac{E_s}{1-\nu_s^2}, \tag{51}$$

By assuming a typical value of 0.33 for $\nu_s$ of globular proteins [53,54], $E^*$ is comparable to $E_s$ as follows:

$$E^* \approx 1.122 \cdot E_s, \tag{52}$$

We first examine the distance dependence of Eq. (49), which exhibits that $F_{\text{ts}}$ increases as the distance decreases, with the force gradient scaling proportionally to $E^*$ (Fig. 4(a)). Subsequently, we simulated the equation of motion (Eq. (1)) coupling with Eq. (49) using the velocity Verlet algorithm and typical HS-AFM parameters: $k_{\text{cl}}$ = 0.1 N/m, $Q_{\text{cl}}$ = 1.5, $R_{\text{tip}}$ = 5 nm, and $A_{\text{free}}$ = 3 nm$_{\text{p-0}}$. The average forces were calculated from the time-averaged deflection. In Fig. 4(b,c), when $\tilde{A}_{\text{cl}} > 0.5$, good agreement between the simulation and analytical solutions is obtained at both the MinForce and resonance frequencies for $E^* \geq 100\,\text{MPa}$. In contrast, when $E^* \leq 10\,\text{MPa}$, good agreement is limited to values of $\tilde{A}_{cl} > 0.8$. In nondestructive biomolecular imaging, $\tilde{A}_{cl}$ is typically set to 0.9 or more, indicating that this force conversion method can be adapted to a wide range of systems.

Furthermore, proteins and nucleic acids, which are the most amenable to HS-AFM observation, generally have $E_s$ on the order of GPa. Although lipid membranes and liposomes in a fluid phase have $E_s$ of several MPa, the apparent $E_s$ in AFM experiments is an order of magnitude higher than the actual value due to the bottom effect [51,52]. Therefore, it is rare for the observed target to have $E_s$ of less than 10 MPa.

Note that in Fig. 4(b,c), with $E^* = 1$ GPa, some step-like jumps are observed that are not reproduced in the analytical solution. A frequency spectrum analysis revealed that this discontinuous change in amplitude occurs when $f_0$ is shifted by the interaction and coincides with a harmonic frequency of the excitation frequency.

For the simulation in the attractive regime, we used the Derjaguin, Muller, Toporov, and Maugis (DMT-M) theory [26], where the force $F_{DMT}$ is given by

$$F_{DMT}(z) = \begin{cases} F_{vdW}(z) & \text{for } z \geq z_0, \\ F_{Hertz}(z_0 - z) + F_{vdW}(z_0) & \text{elsewhere,} \end{cases} \qquad (53)$$

where the van der Waals force ($F_{vdW}$) is expressed by

$$F_{vdW}(z) = -\frac{A_H^* R_{tip}}{6z^2}. \qquad (54)$$

The effective Hamaker constant $A_H^*$ is calculated from those of sample $A_H^s$ and tip $A_H^{tip}$ as follows:

$$A_H^* = \sqrt{A_H^s A_H^{tip}}. \qquad (55)$$

Typical values of $A_H$ range between 20 to 500 zJ [55], with hydrocarbons having 20−100 zJ, oxides having 60−200, and metals having 200−500 zJ.

The calculations were conducted with typical ambient AM-AFM parameters, $R_{tip} = 5$ nm, $A_{free} = 10$ nm$_{p-0}$, $E^* = 10$ GPa, and $z_0 = 0.5$ nm. In Fig. 4(d), the calculated force curve showed a gradual increase in the attractive force as the tip approaches the surface, with the magnitude of the attractive force increasing in proportion to $A_H^*$. Subsequently, as the tip enters the Hertzian contact regime, the force exhibits a notable positive increase.

Subsequently, we performed simulations for the attractive regime using Eq. (53), following the repulsive regime. In Fig. 4(e), the simulation and the analytical results for MinForce were in better agreement for all $A_H$. For the result of $f_0$ (Fig. 4(f)), although a similar nonlinear shape is observed, the analytical solution slightly overestimates the force, suggesting that the analytical calculation is valid only for the Maxslope and MinForce frequencies in the attractive regime. Considering the above results, we can conclude that the approximate value of the force can be obtained using this analytical equation for both the repulsive and attractive regimes, particularly when excited at the MaxSlope and MinForce frequencies.

## 5. Experimental Validation of the Theory

To validate that the applicability of the force conversion to practical experiments, where non-ideal effects can occur, we performed force curve measurements using a home-built HS-AFM setup [21], whose operational principles is identical to those of conventional AM-AFM. We used a small cantilever (BLAC10DS-A2, Olympus) with $f_0$ of 682 kHz, $Q_{cl}$ of 1.8, and $k_{cl}$ of 0.13 N/m, which were determined from a thermal spectrum [56]. To compare the static and dynamic forces quantitatively, it is important to consider the dynamic-to-static correction factors $\chi$ and $\xi$ for the optical lever sensitivity and the spring constant, respectively [57]. However, we did not apply any corrections for $k_{cl}$, as these two coefficients are assumed to cancel each other out in our setup, for the following reason. The correction for the spring constant is expressed as $\xi \chi^2$, where the factor $\chi$ takes a fixed value of ~1.03, while $\xi$ depends on the position of the laser spot on the cantilever. Since we used a small cantilever with a length of ~10 μm, which is the same as the laser spot size, the laser was irradiated at the center, resulting in $\xi$ being ~0.97 [58]. By taking these values, $\xi \chi^2$ is calculated to be 1.03, which appears to fall within the experimental error.

The cantilever was dynamically excited using the piezoacoustic method with an $A_{free}$ of 2.8 ± 0.1 nm at various $f_{drive}$ from 217 to 960 kHz, which correspond to $\tilde{\omega}_{drive}$ ranging from 0.32 to 1.41. For the deflection signal acquisition, a fourth-order low-pass filter was used to prevent the oscillation signal from entering the AD converter. Mica was used as the substrate due to its relatively high $E_s$ and its common use as a supporting substrate for biomolecules. The tip was brought close to the sample surface at a constant velocity of 1.5 μm/s and returned to the farthest position when the deflection reaches a threshold value of 2.7 nm, corresponding to an average force of 350 pN. To suppress instrumental noise, the averages of 300 successively measured curves were calculated.

The correlation between the force and amplitude is examined in Fig. 5(a), which demonstrated that as the amplitude decreased, the force tended to increase, as theoretically predicted. Below $f_{peak}$, the force slope at $\tilde{A}_{cl} \sim 1$ became linear and varied as a function of $f_{drive}$. In contrast, above $f_{peak}$, the force slope increased markedly, resulting in a nonlinear curve. Furthermore, as $\tilde{\omega}_{drive}$ reached 1.2, $\tilde{A}_{cl}$ became greater than 1, indicating a two-valued curve, as predicted by the theory. Note that the original force curve data have been analyzed elsewhere [59].

To validate quantitative agreement between the experiments and theory, we performed analytical calculations and simulations under the same conditions as the experiments. $E^*$ was set at 800 MPa, determined experimentally using static-mode AFM with the same threshold as that used in the dynamic-mode AFM experiments (Supporting Information 4). This value is smaller than the bulk mica value of 60 GPa [60] possibly because the apparent $E_s$ depends on the loading force [61], which was small in this experiment.

The analytical results shown in Fig. 5(b) exhibits a high degree of similarity to those of the experiments conducted below $f_{peak}$. However, above $f_{peak}$, they exhibit a more pronounced two-valued characteristic than the experiments. The simulation results in Fig. 5(c) are almost the same as the analytical results, except that the amplitude and force above $f_{peak}$ are slightly attenuated (Supporting Information 5 for the complete graph). To elucidate this discrepancy, the $\tilde{\omega}_{drive}$ dependence of the force at $\tilde{A}_{cl} = 0.9$ is plotted in Fig. 6. This reveals that all results, including the experimental data, exhibit a concave-shaped curve near the lower MaxSlope and MinForce frequencies as the vertex. The discrepancy between the experiment and the analytical calculation at the force minimum is only ~10%, which falls within the measurement error of AFM. Consequently, it is determined that the analytical calculation provides a sufficient approximation.

However, compared to the theory, the experiment shows that the force at low frequencies is larger, while the force at high frequencies was smaller. The former discrepancy appears to be attributable to cantilever base oscillation, while the latter seems to result from dissipative effects not accounted for in the theory, as they manifest only when the tip makes hard contact with the sample.

To substantiate this hypothesis, we also conducted simulations incorporating dissipative effects using the Kelvin-Voigt model (Supporting Information 5). The simulation results with the viscosity of 30 Pa·s are also displayed in Fig. 6. This shows that the force above $f_0$ is effectively suppressed and is in close agreement with the experimental results. In summary, the analytical solution does not hold above $f_{peak}$; however, it provides a satisfactory approximation when excited at the resonance slope.

# 6. Conclusions

In this study, we succeeded in elucidating a previously unaddressed issue in that the force converted from the amplitude value in AM-AFM is substantially larger than the typical molecular binding forces. By analyzing the equation of motion of the cantilever, we revealed that whereas AM-AFM experiments for fragile molecules are generally performed with excitation of the cantilever at the resonance slope frequencies, the conventional conversion equation is only valid for the excitation at $f_0$. This discrepancy results in a 5-fold overestimation of the actual force. The theoretical formulation shown here is valid for the attractive forces as well as repulsive forces in both ambient and liquid environments. By exciting at the resonance slope, we can not only reduce imaging forces but also estimate quantitative forces through a simple analytical solution. This is expected to facilitate our understanding of force-related material research fields that utilize AM-AFM.

## Figures

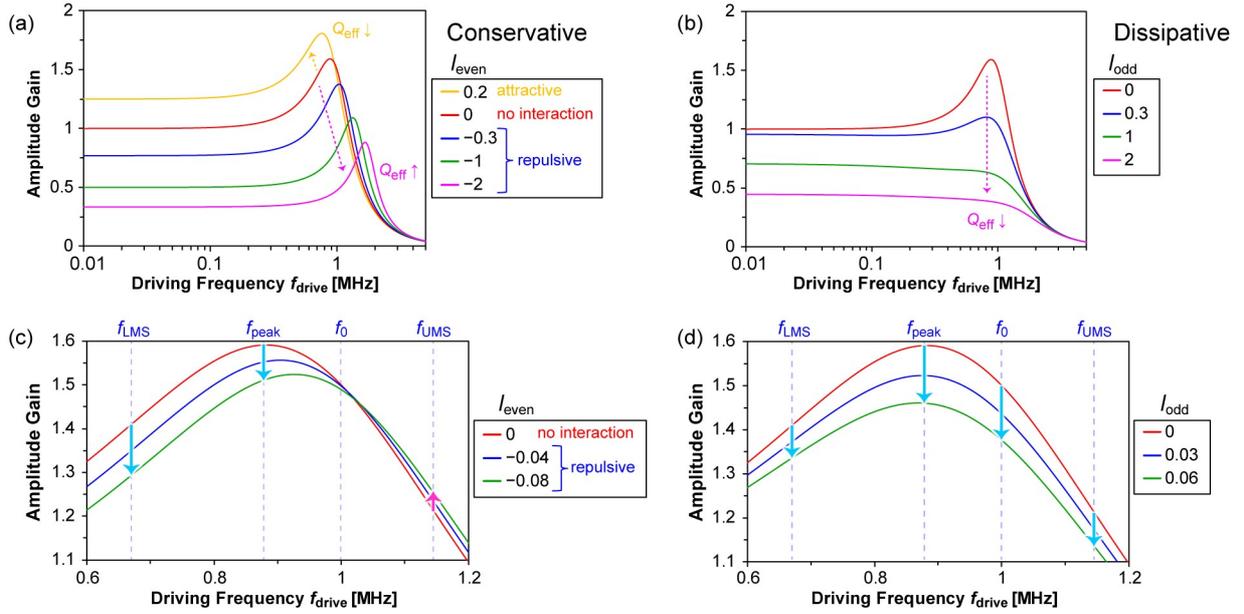

**FIG. 1.** (**a,b**) Theoretical resonance characteristics of the amplitude gain dependent on $I_{even}$ (a) and $I_{odd}$ (b). Positive and negative signs of $I_{even}$ correspond to the attractive and repulsive forces, respectively. (**c,d**) Enlarged views of the resonance amplitude gain dependent on $I_{even}$ (c) and $I_{odd}$ (d). The vertical broken lines indicate the amplitude variation at several characteristic frequencies: $f_{LMS}$, $f_{peak}$, $f_0$, and $f_{UMS}$ represent the lower MaxSlope, peak, resonance, and upper MaxSlope frequencies, with values of 0.67, 0.88, 1, and 1.14 MHz, respectively. All the data are calculated with $Q_{cl} = 1.5$, a typical value for HS-AFM experiments.

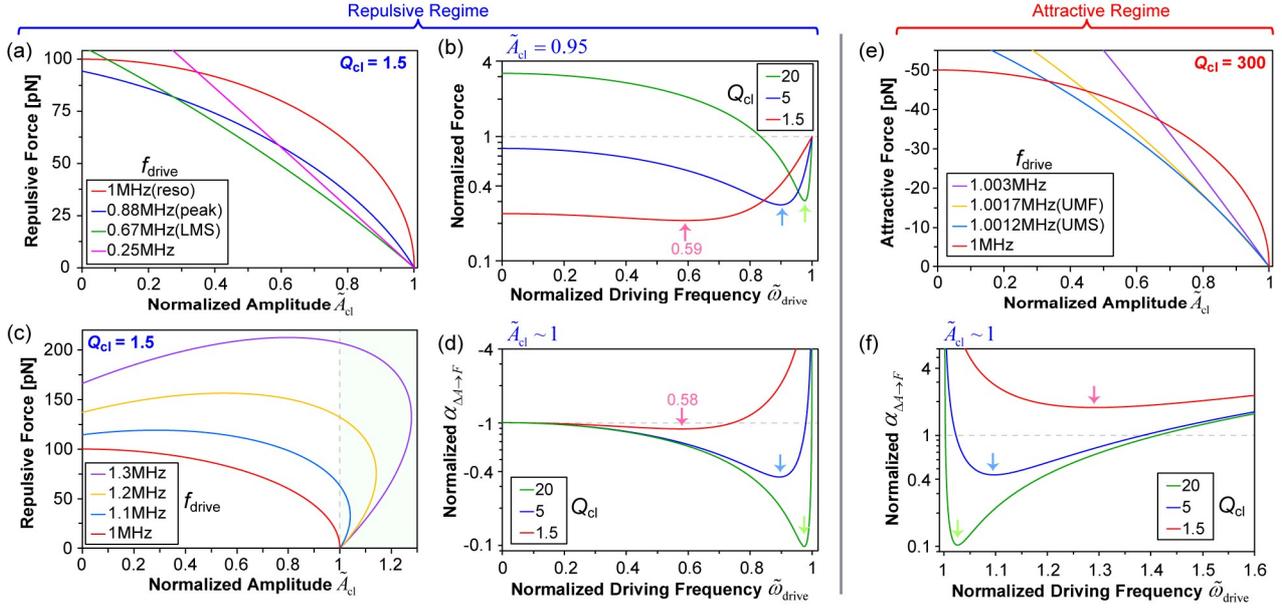

**FIG. 2.** (**a,c**) Theoretical normalized amplitude dependence of the average repulsive force when the driving frequency is below (a) or above (c) the resonance frequency (Eq. (14)). In the legend, "reso", "peak", and "LMS" represent resonance, peak, and lower MaxSlope frequencies, respectively. (**b**) Theoretical driving frequency dependence of the average repulsive force ($\tilde{A}_{cl} = 0.95$) normalized by those at the resonance frequencies. (**d,f**) Driving frequency dependence of $\alpha_{\Delta A \to F}$ ($\tilde{A}_{cl} = 1$) normalized by those at 0 Hz for different $Q_{cl}$ in the repulsive (d) and attractive (f) regimes (Eq. (23)). The arrows in (b,d,f) indicate the force minima for each $Q_{cl}$. (**e**) Normalized amplitude dependence of the average attractive force (Eq. (14)). In the legend, "UMS" and "UMF" represent the upper MaxSlope and MinForce frequencies, respectively.

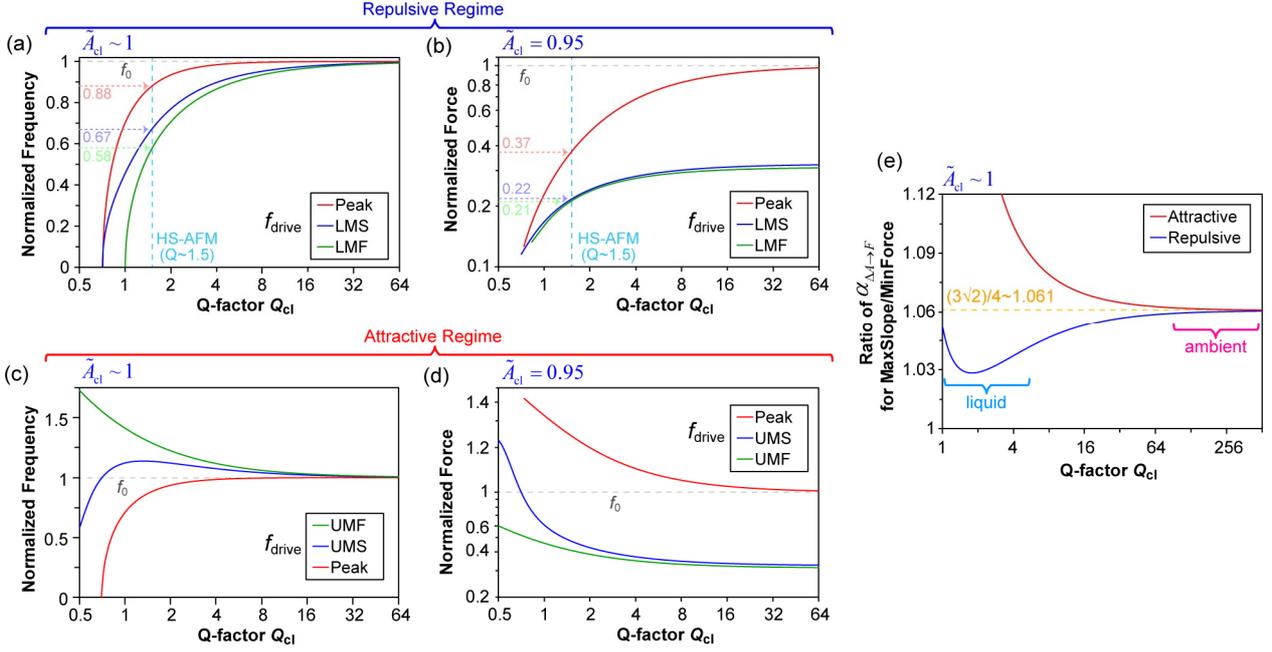

**FIG. 3.** (**a,c**) Theoretical $Q_{cl}$ dependence of characteristic frequencies normalized by the resonance frequencies ($f_0$) in the repulsive (a) and attractive (c) regimes. (**b,d**) Theoretical $Q_{cl}$ dependence of the average force ($\tilde{A}_{cl} = 0.95$) that is normalized by those at the resonance in the repulsive (b) and attractive (d) regimes. In the legend, "LMS", "LMF" represent the lower MaxSlope and MinForce frequencies, respectively, and "UMS", "UMF" represent the upper MaxSlope and MinForce frequencies, respectively. The dashed horizontal lines indicate the results at $f_0$. (**e**) $Q_{cl}$ dependence of the ratio of $\alpha_{\Delta A \to F}$ between the MaxSlope and MinForce frequencies, with the lower and upper frequencies used for the repulsive and attractive regimes, respectively.

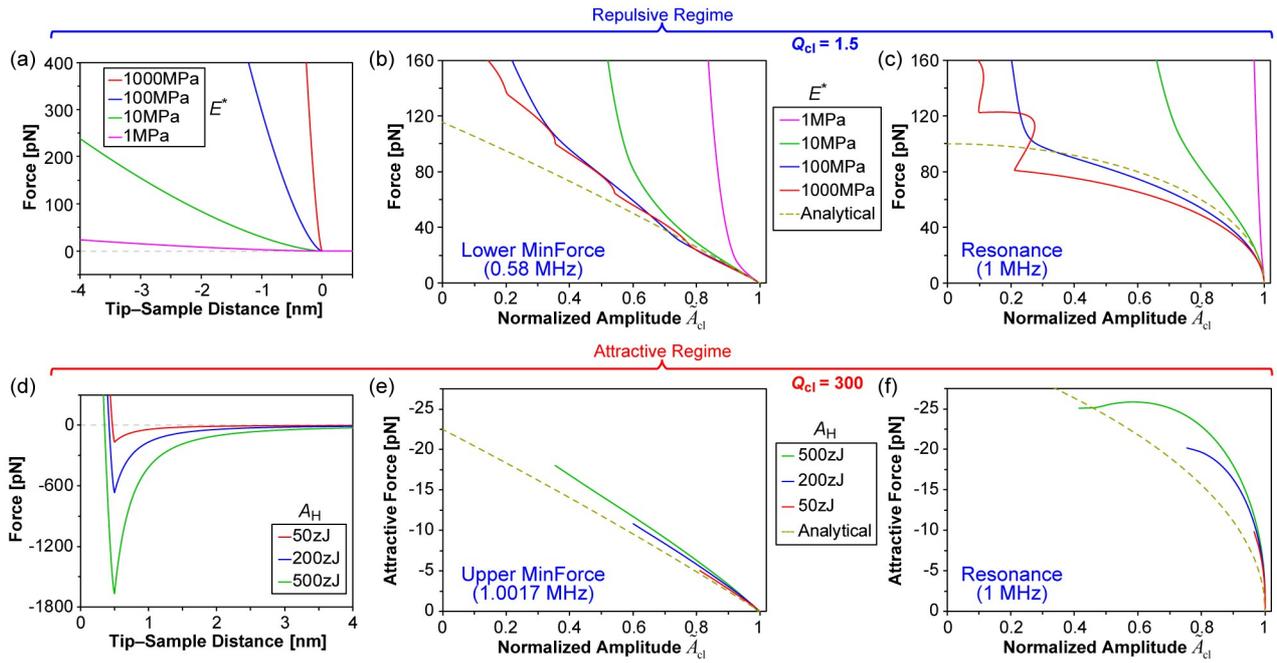

**FIG. 4.** (**a,d**) Force-distance curves used in the simulations in the repulsive (a) and attractive (d) regimes, performed based on the Hertzian and DMT models, respectively. (**b,c,e,f**) Numerical simulation and analytical results of force as a function of normalized amplitude in the repulsive (b,c) and attractive (e,f) regimes when the driving frequency is at the MinForce (b,e) and resonance (c,f) frequencies. The analytical calculations are represented by Eq. (14).

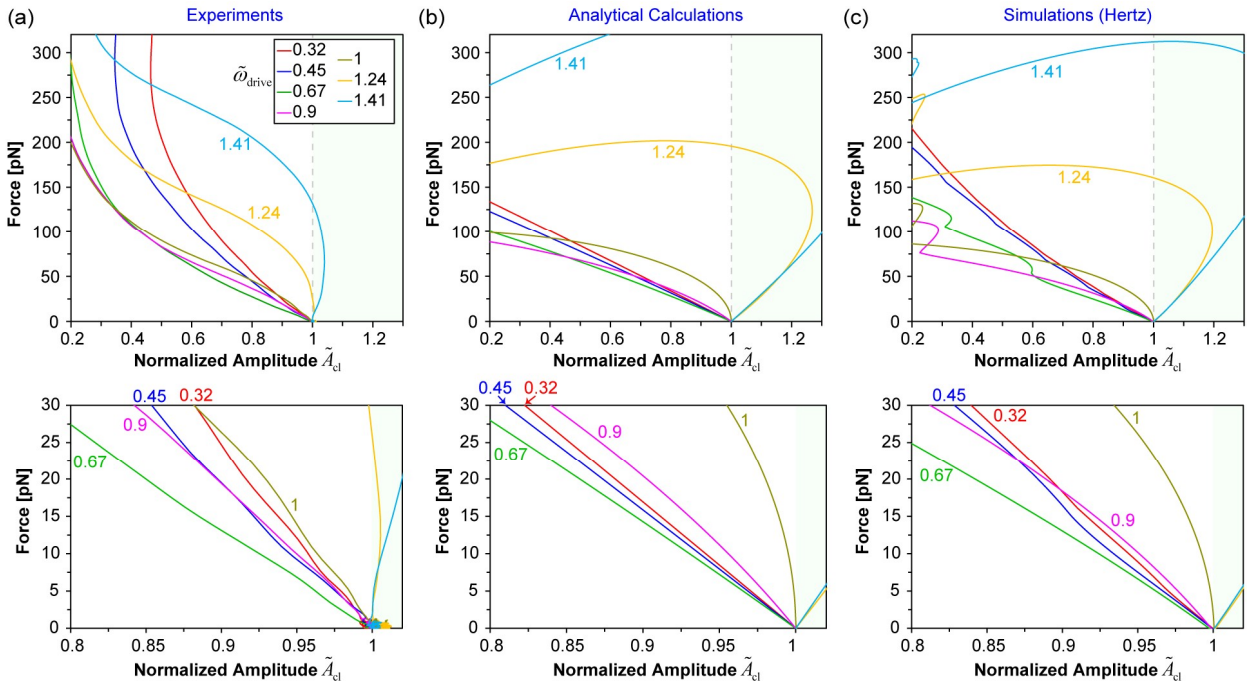

**FIG. 5. (a–c)** Comparison of experiment (a), analytical calculation (b), and simulation (c) of the amplitude dependence of the mean force using various driving frequencies. The magnified graphs near $\tilde{A}_{cl} = 0.9$ are displayed in the lower section. The correspondence between the driving and characteristic frequencies is illustrated in Fig. 6.

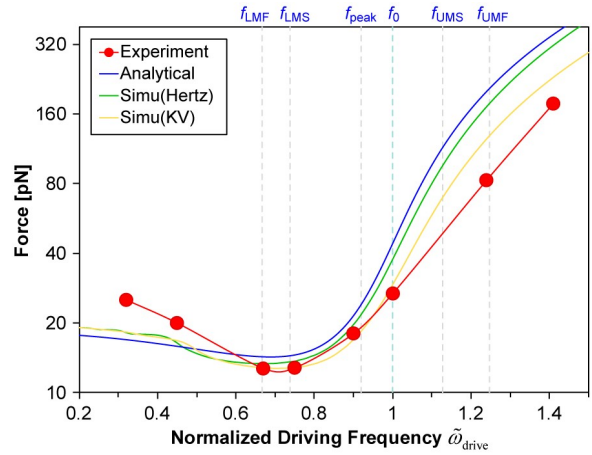

**FIG. 6.** Comparison of experimental data, analytical calculations, and simulations using the Hertzian and Kelvin-Voigt (KV) models for the dependence of average force on driving frequency at $\tilde{A}_{cl} = 0.9$. The standard error of the experimental data, ranging from 0.19 to 0.42 pN, is comparable to the line thickness; thus, error bars are omitted.

**Acknowledgments**

We thank Dr. Steven J. McArthur at NanoLSI, Kanazawa University for critical reading and the English language improvement of the paper. This work was supported by PRESTO, Japan Science and Technology Agency (JST) [JPMJPR20E3 and JPMJPR23J2 to K.U.]; and KAKENHI, Japan Society for the Promotion of Science [21K04849 (to K.U.), 20H00327, and 24H00402 (to N.K.)].

**Author contributions**

K. U. constructed the theories and wrote the manuscript; K. K. supported the study; and N. K. supervised the study.

**Data availability**

The data that support the findings of this study are available from the corresponding author upon reasonable request.